# The Science of al-Biruni

Amelia Carolina Sparavigna
Department of Applied Science and Technology, Politecnico di Torino, Italy

**Abstract:** Al-Biruni (973–1048) was one of the greatest scientists of all times. He was an astronomer, mathematician and philosopher, and studied physics and natural sciences. In this paper, we will discuss some of his experimental methods and some instruments he used.

**1. Introduction**
George Sarton, the founder of the History of Science discipline, defined al-Biruni as "one of the very greatest scientists of Islam, and, all considered, one of the greatest of all times" [1,2]. A universal genius that lived in the Central Asia a thousand of years ago, al-Biruni "was so far ahead of his time that his most brilliant discoveries seemed incomprehensible to most of the scholars of his days", so wrote Bobojan Gafurov in his article on the Unesco Courier [3].
Abū al-Rayḥān Muḥammad ibn Aḥmad al-Bīrūnī (973–1048), was born in Kath, Khwarezm [4]. Khwarezm, also known as Chorasmia, is a large oasis region in western Central Asia, bordered by Aral Sea and deserts. It was the country of the Khwarezmian civilization and of several kingdoms. Today, it is fractioned and belongs to Uzbekistan, Kazakhstan and Turkmenistan. Leaving his homeland, al-Biruni wandered in Persia and Uzbekistan. Then, after Mahmud of Ghazni conquered the emirate of Bukhara, Al-Biruni moved in Ghazni. This town, which is in modern Afghanistan, was at that time the capital of Ghaznavid dynasty [4-6]. In 1017, al-Biruni travelled to the Indian subcontinent, studying the Indian science and conveying it to the Islamic world [4,5].
Al-Biruni was an astronomer, mathematician and philosopher, studying physics and natural sciences too. He was the first able to obtain a simple formula for measuring the Earth's radius. Moreover, he thought possible the Earth to revolve around the Sun and developed the idea the geological eras succeed one another [3]. In fact, in his scientific body of work he addresses almost all the sciences [4,7]. He had excellent knowledge of ancient Greek and studied several works by ancient Greek scientists in their original forms; among them there were the Aristotle's Physics, Metaphysics, De Caelo, and Meteorology, the works of Euclid and Archimedes, the Almagest of the mathematician and astronomer Ptolemy [7,8]. "When religious fanaticism swept medieval Europe… al-Biruni, as a forerunner of the Renaissance, was far in advance of the scientific thought then obtaining in Europe" [7,8]. After a short discussion on his life, let us review some experimental methods and instruments this outstanding man proposed and used.

**2. Life and Works**
As previously told, al-Biruni was born in Kath, a district of Khwarezm. In fact, the word "Biruni" means "from an outer district", in Persian, and so he was known as "the Birunian", with the Latinised name "Alberonius" [4,9]. In his early youth, fortune brought al-Biruni in contact with an educated Greek who was his first teacher [3]. His foster father, Mansur, was a member of the royal family and a distinguished mathematician and astronomer. He introduced al-Biruni to Euclidean Geometry and Ptolemaic astronomy [3]. Then, al-Biruni spent his first twenty-five years in Khwarezm where he studied the body of Islamic law, theology, grammar, mathematics, astronomy and other sciences. In the time, Khwarezm had long been famed for its advance culture. Its cities had magnificent palaces and religious colleges, and the sciences were esteemed and highly developed [3].

Leaving his homeland, al-Biruni wandered, unsettled, for a brief period of time. He was interested in continuing his studies in astronomy, but this would be possible only in a large city. Then, al-Biruni settled on Ravy, which was located near the present day Teheran [10]. Unfortunately, in 996, al-Biruni was not yet well known outside of Kath and then he was unable to find a patron in Ravy; he was poor but remained confident and continued to study [10]. It happened that al-Khujandi (940-1000), a respected astronomer, recorded in 994 the transit of the Sun near the solstices, measuring the latitude of Ravy. Al-Biruni found al-Khujandi's results inaccurate. In his "The Determination of the Coordinate of Locations and for Correctly Ascertaining the Distances between Places", al-Biruni explained that the problem was in the sextant used for measurements. Because of this observation, he began to be accepted by other scholars and scientists [10].

In 998, al-Biruni went to the court of the Amir of Tabaristan [4]. There he wrote an important work, known as the "Chronology of Ancient Nations". Al-Biruni explained that the aim of his work was to establish, as accurately as possible, the time span of various eras [3]. The book is also discussing various calendar systems such as the Arabian, Greek and Persian and several others [3]. When Mahmud of Ghazni conquered the emirate of Bukhara (1017), he took all the scholars to his capital Ghazni. Al-Biruni spent then his life serving Mahmud and later his son Mas'ud. He was the court astronomer and accompanied Mahmud during the invasion of the north-west of India, living there for a few years [4]. During this time, he wrote the "History of India", ending it around 1030. Let us note that most of the works of Al-Biruni are in Arabic although he wrote one of his masterpieces, the Kitab al-Tafhim, both in Persian and Arabic [4].

Al-Biruni catalogued both his own works and those of al-Razi. In 1035-36, or a little thereafter, al-Biruni wrote, at the urging of a friend, an "Epistle Concerning a List of the Books of Mohammad ibn Zakarīyā' al-Rāzī" [11]. This epistle consists of two parts, the first devoted to al-Razi and his works, the second to al-Biruni himself with an inventory. This sort of bibliographical treatment is modelled on those produced by Galen in antiquity [11]. Al-Biruni's catalogue of his own literary production lists 103 titles divided into 12 categories: astronomy, mathematical geography, mathematics, astrological aspects and transits, astronomical instruments, chronology, comets, an untitled category, astrology, anecdotes, religion, and books of which he no longer possesses copies [4,11]. His extant works include the "Indica, a Compendium of Indian Religion and Philosophy", the "Book of Instruction in the Elements of the Art of Astrology", and the abovementioned "Chronology of Ancient Nations". We find also "The Mas'udi Canon", an encyclopaedic work on astronomy, geography and engineering, dedicated to Mas'ud, son of Mahmud of Ghazni, "Understanding Astrology", which is a book containing questions and answers about mathematics and astronomy, the "Pharmacy", about drugs and medicines, "Gems" a book on geology, minerals and gems, dedicated to the son of Mas'ud, the "Astrolabe", the "History of Mahmud of Ghazni and his Father" and the "History of Khwarezm" [4].

**3. Earth, Heaven and Astronomy**
Al-Biruni dealt with Earth in many of his works [12]. He proposed a method to measure its radius, using trigonometric calculations. Let us see how he did. First of all, he measured the high of a hill by measuring the angles subtended by the hill at two points a known distance apart. Then he climbed the hill and measured the angle of the dip of the horizon [13]. In the Figure 1, it is shown the method as discussed in [13].

Using an Arabic mile equal to 1.225947 English miles, al-Biruni value of the radius was equal to 3928.77 English miles, which compares favourably, being different of 2%, with the mean radius of curvature of the reference ellipsoid at the latitude of measurement; this mean radius is of 3847.80 miles [14]. He did this when he was at the Fort of Nandana in Punjab [15]. Since the al-Biruni's self-constructed instrument could have measure angles up to 10' of the arc, the key to the precision of the measurement is a precise sine value, which he seems to have obtained from various Indian sources [14].

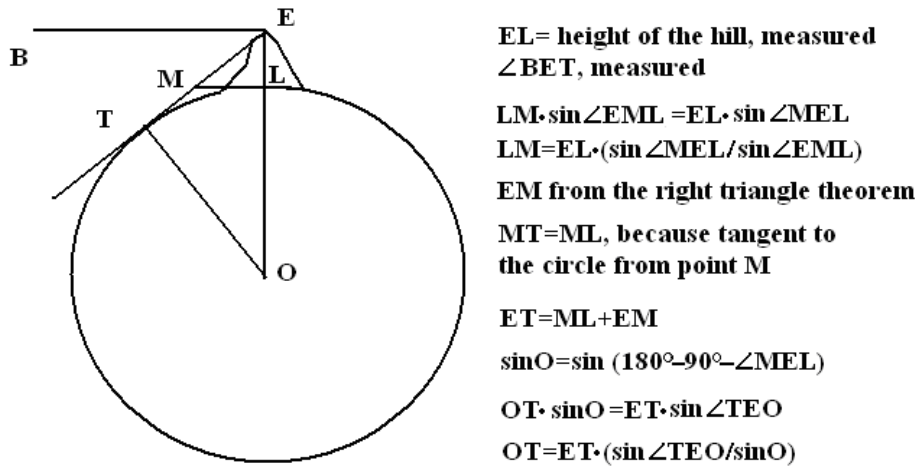

Figure 1 - Al-Biruni's method to measure the radius of the Earth, from Ref.13

As discussed in [12], al-Biruni considered the world, that is the universe, had come into existence in time, as Muslims believed, and then it was not eternal like Aristotle told. However, it is impossible to determine the creation of the world in term of human calculations. The Earth arose from the natural adjustment of the four elements with each other at the centre of the universe, and all the heavenly bodies gravitate towards it. The Earth is a globe, with a rough surface due to the presence of mountains and depressions, but these are negligible when compared with the size of the globe. Because of this irregular surface, the water is not covering it completely, as it would happen for a smooth sphere.
"While water, like earth, has a certain weight and falls as low as possible in the air, it is nevertheless lighter than earth, which therefore settles in water, sinking in the form of sediments at the bottom… The earth and the water form one globe, surrounded on all sides by air. Then, since much of the air is in contact with the sphere of the Moon, it becomes heated in consequence of the movement and friction of the parts in contact. This there is produced fire, which surrounds the air, less in amount in the proximity of the poles owing to the slackening of the movement there" [12]. When discussing the geological changes on the Earth, al-Biruni says that "the center of gravity of the Earth also changes its position according to the position of the shifting matter on its surface" [12]. "With the passing of time, the sea becomes dry land, and dry land the sea" al-Biruni wrote [3], but "if such changes took place on earth before the appearance of man, we are not aware of them" [12]. For instance, he tells of the Arabian desert, which was a sea and then became filled of sand. He also reports of the discovery of "stones which if broken apart, would be found to contain shells, cowry-shells and fish-ears". By "fish-ears" he must have meant fossils [12].
In the Mas'udi Canon, al-Biruni writes that the Earth is at the centre of the universe and that it has no motion of its own, as it is in the Ptolemaic system. However, in this book, he takes issue with this system on several points. "He holds, for example, that the Sun's apogee is not fixed, and while he accepts the geocentric theory, he shows that the astronomical facts can also be explained by assuming the Earth revolves around the Sun" [15]. Then, continuing his speculation on the motion of the Earth, al-Biruni tells that he could neither prove nor disprove it, but commented it favourably [4]. It seems also that he wrote in a commentary on Indian astronomy that he resolved the matter of Earth's motion in a work on astronomy that is no longer extant, his "Key to Astronomy".
Let us summarize his point of view reporting what he tells us about an astronomical instrument, the "Zuraqi", probably an armillary sphere or a spherical astrolabe, or even a mechanical astrolabe. Al-Biruni writes that Sijzi, a Persian astronomer and mathematician from Sistan, a region lying in the south-west of Afghanistan and south-east of Iran, invented an astrolabe the design of which was based on the idea that the Earth moves [4,16,17]: "I have seen the astrolabe called Zuraqi invented by Abu Sa'id Sijzi. I liked it very much and praised him a great deal, as it is based on the idea entertained by some to the effect that the motion we see is due to the Earth's movement and not to that of the sky. By my life, it is a problem difficult of solution and refutation. … For it is the same whether you take it that the Earth is in motion or the sky. For, in both cases, it does not affect the Astronomical Science. It is just for the physicist to see if it is possible to refute it" [4,16].

**4. The Zijes**
The Islamic Golden Age (8th-15th centuries) strongly promoted the astronomy and several scholars contributed to its development. The Islamic scientists assimilated and amalgamated disparate material to create their astronomical science. This material included Greek, Sassanid, and Indian works in particular [18]. In turn, Islamic astronomy had a significant influence on the astronomy of the medieval Europe. Many stars and astronomical terms such as alidade, azimuth, and almucantar, are still referred to by their Arabic names [18]. From 700 to 825, we have the period of assimilation and syncretisation of earlier Hellenistic, Indian, and Sassanid astronomy. Some first astronomical texts, translated into Arabic, had Indian and Persian origin. The most notable of these texts was the "Zij al-Sindhind", an 8th-century Indian astronomical work that was translated by al-Fazari and Yaqub ibn Tariq after 770 under the supervision of an Indian astronomer who visited the court of Abbasid caliph al-Mansur [18]. During this period, the Arabs adopted the sine function, inherited from Indian geometry, instead of chords of arc used in Greek trigonometry [18,19]. From 825 to 1025, there was a period of vigorous investigation, in which the Ptolemaic system of astronomy was accepted, however, under the possibility of observational refinements and mathematical revisions [18,19]. One of the major works was the "Zij al-Sindh" written by al-Khwarizmi in 830. In this period, a great impulse to astronomical research came from the Abbasid caliphs. They supported this scientific work financially and gave it a formal prestige [18].
Zij is the generic name of Islamic astronomical books that tabulate parameters used for astronomical calculations concerning the positions of the Sun, Moon, stars, and planets. The name is derived from a Persian term meaning cord. May be, this is a reference to the arrangement of the threads on a loom, like the tabulated data are arranged in rows and columns [20]. Let us remark that the medieval Muslim zijes were more extensive, typically including materials on chronology, and the geographical latitudes and longitudes. Going beyond the traditional contents, some zijes even explain the theory or report the observations from which the tables were computed [20]. Besides the Zij written by al-Khwarizmi, other famous zijes are those of the Egyptian astronomer Ibn Yunus (c. 950-1009). In one of them he described, with precision, forty planetary conjunctions and thirty lunar eclipses [21]. His astronomical tables give data obtained with very large astronomical instruments and the use of trigonometric identities [22].
Probably it was not the entire driving force to this growth of astronomy, but religion contributed to it [21]. In fact, the Islam needed a way to figure out how to orient all sacred structures toward Mecca [21]. And then a precise celestial mapping was necessary to find the right direction, or qibla, toward the Kaaba. By the 9th century, the astronomers were commonly using trigonometry to determine the qibla from geographical coordinates, turning the qibla determination into a problem of spherical astronomy. Al-Biruni for example, in "The Determination of the Coordinate of Locations and for Correctly Ascertaining the Distances between Places", has the goal to find the qibla at Ghazni.
One of the al-Biruni zijes contains a table giving the coordinates of six hundred places, almost all of them measured by al-Biruni himself. For some places he is reporting data taken from similar tables given by al-Khwarizmi. Al-Biruni seems to have realized that for places given by both al-Khwarizmi and Ptolemy, the value obtained by al-Khwarizmi was more accurate [19,21]. Muhammad ibn Mūsā al-Khwārizmī (c. 780 – c. 850) was a Khwarezmian too. In the early 9th century, he produced accurate sine and cosine tables, and the first table of tangents. He was also a pioneer in spherical trigonometry. By the 10th century, Muslim mathematicians were using all six trigonometric functions. Let us note that the term "algorithm" is coming from medieval Latin "algorismus", a mangled transliteration of Arabic al-Khwarizmi, "native of Khwarezm". The earlier form of this word in Middle English was "algorism" (early 13[th] c.) [23].

**5. Quadrants, Astrolabes and Clocks**
As told in [15], al-Biruni was among those deported in Afghanistan by Mahmud of Ghazni . He was then 44 years old. On 14 October 1018, we find him in a village south of Kabul, where he wanted to measure the height of the sun but had no instrument to hand. So he was obliged to draw a calibrated arc on the back of a reckoning board and used it, with the aid of a plumb line, as a makeshift quadrant. On the basis of the measurements made with this crude device he calculated the latitude of the locality. This quadrant was probably an inclinometer based on quarter-circle panel.

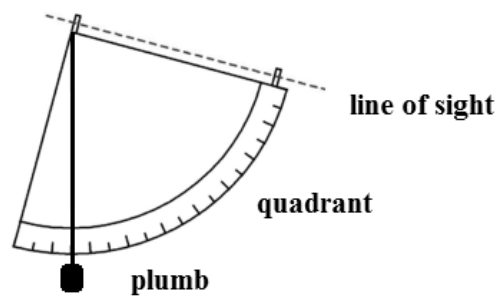

**Figure 2 – A quadrant**.

Along one edge there were two sights forming an alidade. A plumb bob was suspended by a line from the centre of the arc as in the Figure 2. In order to measure the altitude of a star, the observer would view the star through the sights (pinholes in the case of the Sun) and hold the quadrant vertical. The plumb indicates the reading on the graduation. It is better to have a person concentrated on observing the star and holding the instrument and another person to take the reading. The accuracy of such an instrument is limited by its size.
An astrolabe is a more elaborate instrument. It helps in measuring the positions of Sun, Moon, planets, and stars, and it is therefore fundamental to determine the local time at a given latitude and vice-versa. An astrolabe consists of a disk, the "mater", deep enough to hold one or more flat plates called "tympans" [24]. Each tympan is made for a specific latitude and engraved with a stereographic projection of circles denoting azimuth and altitude, and representing the portion of the celestial sphere above the local horizon (see the Figure 3). Two other sets of curves represent the unequal hours and the houses of the heaven. The rim is typically graduated into hours of time, degrees of arc, or both. Above the mater and tympan, there is the "rete", a framework bearing a projection of the ecliptic plane and several pointers indicating the positions of the brightest stars [24]. The rete is free to rotate. When it is rotated, the stars and the ecliptic move over the projection of the coordinates on the tympan. One complete rotation corresponds to a day. On the back of the mater, there is often engraved a number of scales, useful in various applications, and a graduation of 360 degrees around the rim. The alidade is attached to the back face. When the astrolabe is held vertically, the alidade can be rotated and the Sun or a star sighted along its length, so that its altitude in degrees can be read from the graduated edge of the astrolabe [24].

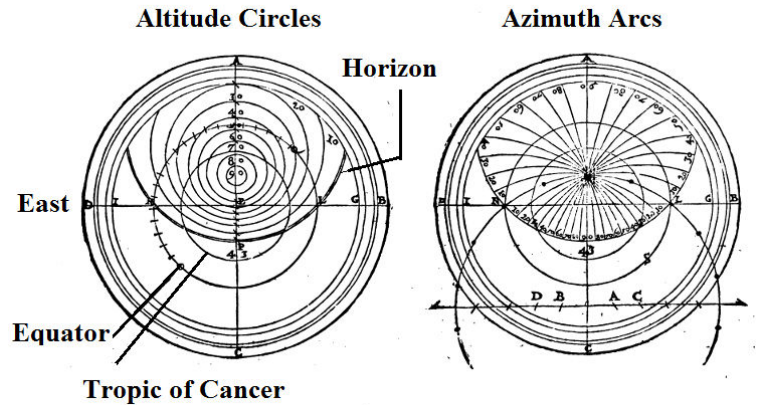

**Figure 3** - Curves of altitude (almucantar) and azimuth on the astrolabe, from the book entitled "Dell'Uso et Fabbrica dell'Astrolabio", by Egnatio Danti, Giunti, Firenze, 1578 [25].

Al-Biruni, in a treatise on the Astrolabe, describes how to tell the time during the day or night and use it, as it can be used a quadrant, for surveying. In fact, the astrolabe is a complex instrument, and all its features have been added over centuries. Moreover, several other instruments have been used at the time of al-Biruni. Reference 26 contains the critical edition with English translation of an Arabic treatise on the construction of over one hundred various astronomical instruments, composed in Cairo ca. 1330, with citations to the al-Biruni works.

The mechanical astrolabes with gears were invented in the Muslim world. These geared instruments were designed to produce a continual display of the current position of Sun and planets. We find a device with eight gear-wheels (Figure 4, on the right) illustrated by al-Biruni in 996, so that this al-Biruni mechanism can be considered an ancestor of the astrolabes and clocks developed by later Muslim engineers. The same author of [26], François Charette, is considering it a simpler version of the Antikythera mechanism [27], such as previously proposed by Derek J. de Solla Price [28].

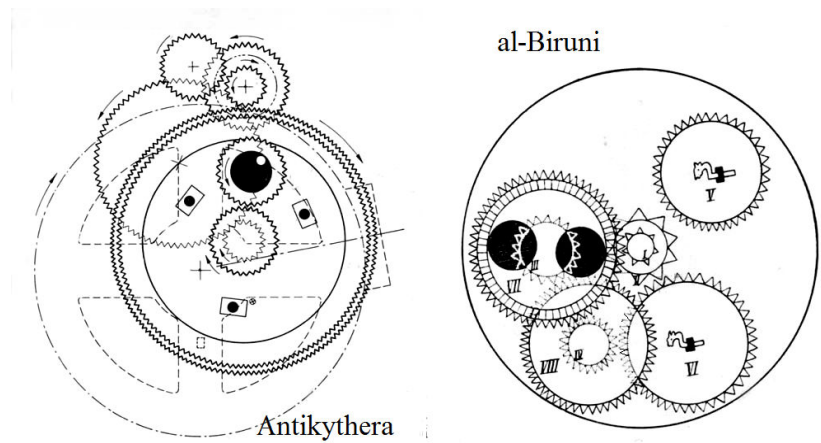

**Figure 4** - On the left, an attempt of reconstruction made by the Rear Admiral Jean Theophanidis [29] of the Antikythera mechanism and, on the right, the al-Biruni mechanism, adapted from Ref.28.

In 1900, a Greek sponge diver discovered the wreck of an ancient ship off the Antikythera island in the Dodecanese. Divers find several bronze and marble statues and other artifacts from the site.
In 1902, an archaeologist noticed that a piece of rock recovered from the site had a gear wheel embedded in it. This rock revealed itself as one of the oldest known geared devices, able to display the motions of Sun, Moon and planets. After decades of work on it, de Solla Price, discussed this mechanism in an article entitled "An Ancient Greek Computer" in the Scientific American of June 1959. He saw a direct connection between devices like the Antikythera machine and the Islamic astrolabes. Several years after, a Byzantine device dating from the 6$^{th}$ century, which models the motions of the Sun and Moon, had been discovered: this device can be used as a link between the Antikythera mechanism and the mechanical instrument described by al-Biruni [30]. It is probable that the Antikythera mechanism was not the only one. Cicero, in the 1$^{st}$ century BC, is mentioning an instrument constructed by the philosopher Posidonius, "which at each revolution reproduces the same motions of the sun, the moon and the five wandering stars (the planets) that takes place in heaven day and night" [30].

**6. A Balance of Wisdom**
Al-Biruni developed experimental methods to determine the density of substance, some based on the theory of balances and weighing and others based on the volume of fluids. He also generalizes the theory of the centre of gravity and applies it to the volumes. As told in [31], "using a whole body of mathematical methods … , Arabic scientists raised statics to a new, higher level. The classical results of Archimedes in the theory of the centre of gravity were generalized and applied to three-dimensional bodies, the theory of ponderable lever was founded and the 'science of gravity' was created and later further developed in medieval Europe. The phenomena of statics were studied by using the dynamic approach so that two trends – statics and dynamics – turned out to be inter-related within a single science, mechanics. … Numerous fine experimental methods were developed for determining the specific weight, which were based, in particular, on the theory of balances and weighing. The classical works of al-Biruni and al-Khazini can by right be considered as the beginning of the application of experimental methods in medieval science".
As told in [32], al-Khāzini (Abu al-Fath Khāzini, who fourished 1115–1130) described an istrument used by al-Biruni in measuring densities. It was a hydrostatic balance. The scales were used to test the purity of metals and to ascertain the composition of alloys. The Arabs used a method based on comparison of the weights of equal volumes: Al-Biruni for example, takes hemispheres of the different metals or rods of equal size and compares their weight [32].

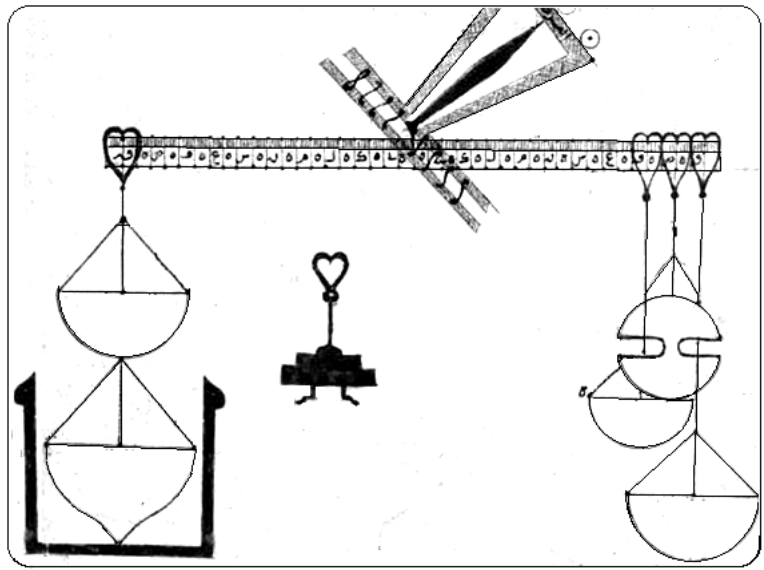

**Figure 5 - A mizan al-hikma, a balance of wisdom, which is in fact a hydrostatic balance, like that of the "The Book of the Balance of Wisdom" by Al-Khāzini.**

In the Figure 5 we can see a drawing of a mizan al-hikma, a balance of wisdom, which is in fact a hydrostatic balance, created after an image from the book of Abu al-Fath Khāzini (flourished 1115–1130), entitled "The Book of the Balance of Wisdom" [33]. Reference [34] tells that, as early as 1857, the year in which the American Oriental Society published in its journal the contribution of

N. Khanikoff on this book, it was known that as far as the determination of the specific gravity, Al-Khazini had drawn much from the work of Al-Biruni.
The hydrostatic balance is an old instrument. The Latin poem "Carmen de Ponderibus et Mensuris" of the 4$^{th}$ or 5$^{th}$ century describes the use of it referring to Archimedes [35,36]. This balance is also linked to a widely known anecdote. A votive crown for a temple had been made for King Hiero II of Syracuse, who supplied the pure gold, and Archimedes was asked to determine whether some silver had been substituted by the goldsmith. Archimedes had to solve the problem without damaging the crown, so he could not melt it down into a regularly shaped body and calculate its density from weight and volume. Concerning the anecdote of the golden crown, Galileo Galilei suggested that Archimedes used the hydrostatic balance.

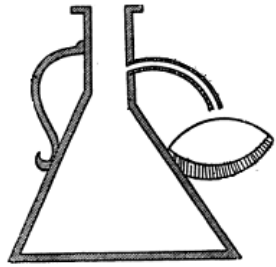

**Figure 6 - The cone-shaped vessel in the Ref.34.**

**7. Vitruvius' and al-Biruni's methods**
However, to evaluate the density or specific weight of materials, al-Biruni refers to another method too. This method is based on the volumes of fluids and on the use of a specific instrument. It was a vessel in which the level of water or oil remained constant, since any excess was drained out of the holes made for this purpose. He was able to measure the displaced water with such exactitude that his findings nearly correspond with modern values [32,34]. The Figure 6 shows this vessel depicted by al-Khāzini, as a cone-shaped vessel. To measure the specific gravities of gemstones, al-Biruni used it.
Before discussing the method, let us read what Vitruvius is writing in his De Architectura, in the chapter entitled "of the Method of Detecting Silver when Mixed with Gold" [37]. "Charged with this commission (to determine whether the crown had silver inside or not), he (Archimedes) by chance went to a bath, and being in the vessel, perceived that, as his body became immersed, the water ran out of the vessel. Whence, catching at the method to be adopted for the solution of the proposition, he immediately followed it up, leapt out of the vessel in joy, and, returning home naked, cried out with a loud voice that he had found that of which he was in search, for he continued exclaiming, in Greek, Eureka, (I have found it out). After this, he is said to have taken two masses, each of a weight equal to that of the crown, one of them of gold and the other of silver. Having prepared them, he filled a large vase with water up to the brim, wherein he placed the mass of silver, which caused as much water to run out as was equal to the bulk thereof. The mass being then taken out, he poured in by measure as much water as was required to fill the vase once more to the brim. By these means he found what quantity of water was equal to a certain weight of silver. He then placed the mass of gold in the vessel, and, on taking it out, found that the water which ran over was lessened, because, as the magnitude of the gold mass was smaller than that containing the same weight of silver. After again filling the vase by measure, he put the crown itself in, and discovered that more water ran over then than with the mass of gold that was equal to it in weight; and thus, from the superfluous quantity of water carried over the brim by the immersion of the crown, more than that displaced by the mass, he found, by calculation, the quantity of silver mixed with the gold, and made manifest the fraud of the manufacturer." What Vitruvius describes is the Archimedean displacing volume method. In Reference [37], I proposed that Archimedes could have used the vessel of a water-clock, that is, of a clepsydra. Moreover, I repeated the experiment to show in detail the method.

Probably al-Biruni read a different report, from a Greek source of this episode. Let us see how al-Biruni could have interpreted it, by describing the method he used to determine the density of a substance. Al-Biruni filled with water the vessel in the Figure 6 until the water began to run out by a pipe at the side; then a definite mass, as large as possible, of the substance is weighed ($P_1$) and the pan ($P_2$) of a scale placed under the outlet pipe [32]. Then, the substance is put in the vessel. This body displaces the water so that it flows in the pan. The pan and the water are weighed ($P_2+P_3$). The difference (($P_3+P_2$)−$P_2$) is the weight of the displaced water. By the ratio $P_1/P_3$ we can have the density of the substance.

Al-Biruni applied the method to determine the density of precious stones. For instance, the sapphire has a specific gravity (the ratio of the density of a substance to the density of a reference substance) of 3.95–4.03, whereas the glass of 2.4–2.8. Using his method, it is possible to distinguish them. For what concerns the accuracy of the method, al-Khāzini remarks that it is difficult to weigh the amount of water displaced, because the water sticks to the sides of the outlet-tubes [34]. And in fact, al-Biruni tells that it is better to use a mass as large as possible in order to increase the accuracy.

The determination of specific gravity played a quite important role in the al-Biruni's researches, and the results he obtained were propagated by various scholars of the Islamic countries. One may ask why this research was so relevant [34]: because al-Biruni acknowledged a social importance for it, that is, an intrinsic worth in metals and jewels. Therefore, certain physical properties had to be found to evaluate them [34]. For instance, al-Biruni objected against the classification of gems on the basis of their colours only, as was the common practice of the time. The colour is a secondary property: specific gravity brilliance and hardness are the relevant properties of materials. The hardness was determined by the use a tip of a sample material and by observing the indentation it is producing [34].

**8. Heat and Light**

Reference [34] is pointing out that, contrary to his astronomy or astrology works, on which he wrote separate treatises, there does not exist a single book devoted exclusively on physics, but it is necessary to read all the books to evaluate his physical researches. And then in [34], after such a research, we find what al-Biruni thought on heat and light.

Aristotle considered heat to be a fundamental quality of the element fire and inherent in all things. There are two types of heat, by which the bodies can be heated: internal or external. Starting from the Aristotle's works, Al-Biruni came to the conclusion that "heat is nothing but the rays of the Sun detached from the body of the Sun towards the Earth" [34]. And then, "the heat exists in the rays, it is inherent in them". As observed in [34], the natural conclusion would be that air is heated by the Sun, but al-Biruni tells that "the warmth of the air is the result of the friction and violent contact between the sphere, moving rapidly, and his body". This is an Aristotelian manner of thinking. In any case, al-Biruni had the merit of understanding the connection between motion and heat, the same we find in the Kinetic Theory of heat [34].

And heat and rays were the subjects of several letters of a correspondence between al-Biruni and Ibn Sīnā, Avicenna, and there we find that the heat is generated by the motion and cold by the rest, and for this reason, the Earth is hot at the Equator and cold at the Poles. Another important discussion between the two scientist was on the propagation of heat and rays of Sun. Al-Biruni's opinion was that that light and heat are immaterial, and that the heat exists in the rays and it is inherent in them. How is therefore the propagation of heat? After this al-Biruni's question, Avicenna answered that the heat was not propagating by itself, but the rays of the Sun are propagating, and the heat is carried by them, like a man in a boat, which is not moving, but his boat is moving [34]. A very interesting discussion between two outstanding persons.

This problem of the propagation of heat leads al-Biruni to study the problem of the nature and propagation of light. He stated that "there is a different opinion regarding the motion of the rays.

Some say, this motion is timeless, since the rays are not bodies. Others say, this motion proceeds in very short time: that, however, there is nothing more rapid in existence, by which you might measure the degree of its rapidity, e.g. the motion of the sound in the air is not so fast as the motion of the rays, therefore the former has been compared with the latter and thereby its time (the degree of its rapidity) has been determined" [34]. According to [34], this is the first reference to the problem of measuring the speed of light.

**9. Al-Biruni's Wisdom**
Let me conclude this paper with some words written by al-Biruni [39], which illustrate quite well the wisdom of this person and his passion for scientific research. It is the parable of the four pupils, from his "Indica".
A man is travelling together with his pupils from some business towards the end of the night. There appears something standing erect before them on the road, the nature of which is impossible to recognize because of darkness. The man turns towards his pupils and asks them what it is. The first says "I do not know what it is", the second "I do not known, and I have no means of learning what it is", the third "It is useless to examine what it is, for the raising of the day will reveal it". It is clear that none of them had attained the knowledge: the first because of his ignorance, the second was incapable and had no means of knowledge by learning, and the third because he was indolent and acquiesced on his ignorance. The fourth pupil did not give an answer: he stood still and then he went on in the direction of the object. On coming near, he found that it was pumpkins on which there was something entangled. He considered that no living man, endowed with free will, could stand still in this situation, and therefore it was a lifeless object. To be sure, he went quite close to it and struck again it with his foot till it fell to the ground. Thus, removed all doubt, he returned to his master and gave him the exact account.